\documentclass[9pt,twocolumn,twoside]{pnas-new}

\usepackage{graphicx}
\usepackage{amsmath,amssymb}

\usepackage{siunitx}

\usepackage[version=4]{mhchem}
\usepackage{chemformula}

\usepackage{newfloat}

\DeclareFloatingEnvironment[
    fileext=los,
    listname={List of Schemes},
    name=Scheme,
    placement=tbp
]{scheme}

\providecommand{\icarus}{Icarus}
\providecommand{\gca}{Geochimica et Cosmochimica Acta}
\providecommand{\apj}{The Astrophysical Journal}
\providecommand{\apjl}{The Astrophysical Journal Letters}

\providecommand{\aap}{Astronomy \& Astrophysics}

\providecommand{\psj}{The Planetary Science Journal}
\providecommand{\nat}{Nature}
\providecommand{\aj}{The Astronomical Journal}

\newcommand{\pKa}{\mathrm{p}K_{\mathrm{a}}}

\newcommand{\der}{\mathrm{d}}

\templatetype{pnasresearcharticle} 

\begin{document}
\pagestyle{plain}

\title{New Hydrolysis Rate Constants Reveal Longest Cyanide Persistence in Cool, Neutral Waters}


\author[a,1]{Sai Shruthi Murali}
\author[a]{Skyla B. White}
\author[a,2]{Paul B. Rimmer}

\affil[a]{Cavendish Laboratory, University of Cambridge, JJ Thomson Avenue, Cambridge CB3 0HE, U.K.}
\affil[1]{ssm54@cam.ac.uk}
\affil[2]{pbr27@cam.ac.uk}

\leadauthor{Sai Shruthi Murali}

\significancestatement{Hydrogen cyanide is a central feedstock in many proposed pathways to the origin of life, but the amount that can accumulate in natural waters depends on how rapidly it is destroyed. We measure one such destruction mechanism-- hydrogen cyanide hydrolysis --across a range of temperatures, pH values, and dissolved salt types, explicitly quantifying the uncertainties in the resulting rate constants. Our results suggest that cyanide persists longest in cold, near-neutral water and further allow us to estimate the concentrations attainable under plausible planetary conditions. More broadly, our results demonstrate the importance of propagating uncertainties in aqueous rate constants through kinetic models to constrain the range of environments consistent with prebiotic scenarios, rather than rejecting them on the basis of a single nominal value. By rigorously characterizing the uncertainty in the hydrolysis rate constant, this work provides a step toward that goal.}

\authorcontributions{Author contributions: S.S.M performed experimental work and initial data analysis. S.B.W carried out the SOUP modeling. PBR contributed to the modeling and supervised both S.S.M and S.B.W. All the authors contributed to writing the manuscript.}
\authordeclaration{The authors declare no competing interest.}
\correspondingauthor{\textsuperscript{1}To whom correspondence should be addressed. E-mail: ssm54@cam.ac.uk}

\keywords{Hydrolysis $|$ Prebiotic chemistry $|$ Cyanosulfidic chemistry $|$ Chemical kinetics $|$ Lifetime $|$}

\begin{abstract}
	Hydrogen cyanide (\ce{HCN}) is a key molecule in prebiotic chemistry, and its availability in water is limited by hydrolysis. In this work, we estimate the hydrolysis rates of \ce{HCN}, particularly the errors associated with the rates as a function of temperature, pH and in the presence of salts containing sulfite, sulfide and phosphate. For pure water, we find an acid-catalyzed hydrolysis rate constant at ${\rm 0 ^{\circ}C}$ of $\ln (k^+_{273}/1\,{\rm M^{-1} s^{-1}}) = -12.7 \pm 1.12$ with an activation energy of $67.1 \, {\rm kJ \, mol^{-1}}$. We find a base-catalyzed rate constant of $\ln (k_{273}^-/1\,{\rm M^{-1} s^{-1}}) = -9.0 \pm 1.27$ at ${\rm 0^{\circ}C}$ with an activation energy of $89.5 \, {\rm kJ \, mol^{-1}}$. These values are consistent with estimates from the literature within our uncertainties at ${\rm pH} > 8$ but diverge from literature values at lower pH. In the presence of salts, hydrolysis is accelerated under acidic conditions to 6--14$\times$ the acid-catalyzed rate without salts. However, at ${\rm pH \gtrsim 8}$ hydrolysis rates become $2.5\times$ times slower with sulfite and sulfide. These results demonstrate the importance of estimating uncertainties associated with rates when constraining the maximum concentrations of prebiotic molecules attainable in natural waters on the Earth and other planets.
\end{abstract}

\doi{\url{www.pnas.org/cgi/doi/10.1073/pnas.XXXXXXXXXX}}

\maketitle
\thispagestyle{firststyle}
\ifthenelse{\boolean{shortarticle}}{\ifthenelse{\boolean{singlecolumn}}{\abscontentformatted}{\abscontent}}{}

\Firstpage




\noindent Hydrogen cyanide (\ce{HCN}) is considered a key molecule in many prebiotic chemical scenarios \cite{Oro1960,Benner2014,Sutherland2016,Sutherland2017,Becker2019}, as its high-energy nitrile bond and capacity to deliver both carbon and nitrogen within a single reactive molecule make it a versatile precursor for the abiotic synthesis of more complex species. \cite{Oro1960,Ferus2015,Saladino2015,Benner2014}.  The availability of hydrogen cyanide on the Hadean Earth, more than 4 billion years ago, depends on a balance between natural sources of \ce{HCN} and its destruction. Proposed sources of \ce{HCN} include impact delivery \cite{Todd2020,McDonald2025}, impact synthesis \cite{Ferus2020}, volcanic degassing \cite{Rimmer2019a,Rimmer2024}, synthesis through photochemical reactions \cite{Tian2011,Rimmer2019b}, and interactions with solar energetic particles \cite{Airapetian2016}. HCN is generated more efficiently under reducing atmospheric conditions \cite{Rimmer2019b}, such as those that may arise within a post-impact atmosphere \cite{Zahnle2020,Wogan2023}. It has been predicted to be present in a wide range of geochemically plausible environments on early Earth or Mars \cite{Rimmer2021}, as well as on icy moons such as Enceladus \cite{Peter2024}, and in the atmospheres of exoplanets \cite{Giacobbe2021,Claringbold2023}. Once formed, \ce{HCN} can be stored as iron–cyanide complexes and subsequently transformed in prebiotically productive ways \cite{Keefe1996,Green2021}. Given its proposed centrality to prebiotic chemistry, understanding the availability of \ce{HCN} on the early Earth represents a crucial link between the origins of life and its planetary context \cite{Ranjan2016,Sasselov2020}.

Despite the low boiling point of pure \ce{HCN}, it is possible to concentrate \ce{HCN} in water at any pH, owing to its reasonably high Henry’s law constant of $10 \, {\rm M \,bar^{-1}}$ at room temperature \cite{Ma2010}. This effect is most pronounced in alkaline waters, where pH at or above the $\pKa$ of \ce{HCN}, 9.2 at room temperature \cite{Verhoeven1990}, favors deprotonation to the non-volatile cyanide anion limiting the loss of \ce{HCN}, via volatilization. However, even with a relatively high Henry’s law constant, low atmospheric partial pressures of \ce{HCN} can result in low equilibrium aqueous concentrations, especially in neutral to acidic waters. The maximum lifetime of cyanide in water is limited by acid-catalyzed and base-catalyzed hydrolysis, however, in prebiotic environments it is also depleted through reactions forming sugars and amino acids, reducing its effective lifetime \cite{Walton2026}. \Endparasplit 

\noindent Stanley Miller was the first to recognize that measuring the hydrolysis rates of \ce{HCN} is critical for understanding the plausibility of prebiotic chemistry on the primitive Earth \cite{Miyakawa2002}. 

Cyanide hydrolysis yields formamide, a useful solvent for many prebiotic chemical reactions \cite{Ochiai1968,Philipp1977,Saladino2012}, which itself hydrolyzes to form ammonium formate. This makes ammonium formate an excellent proxy for the hydrolysis of \ce{HCN}, disentangling hydrolysis from other possible reactions that can destroy \ce{HCN} (See Scheme \ref{sch:hydrolysis}). 


\begin{scheme}[t]
    \centering
            \centering
            \includegraphics[width=0.9\linewidth]{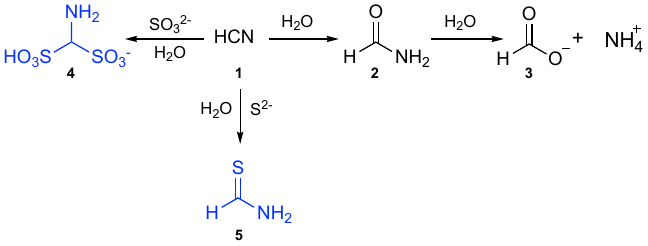}
    \caption{Proposed pathways for the hydrolysis of \ce{HCN} under acidic,
    neutral, and alkaline conditions. The scheme illustrates the hydrolysis of  \ce{HCN} to formate (3) via  the formamide (2) intermediate. In the presence of sulfite and sulfide salts, cyanide forms the intermediate  aminomethanedisulfonate (4) and thioformamide (5) respectively.}
    \label{sch:hydrolysis}
\end{scheme}

The rate of formamide hydrolysis has been found to be 100 times faster than that of \ce{HCN} hydrolysis hindering its accumulation \cite{Miyakawa2002}. This highlights the importance of reaction kinetics in determining the stability of prebiotic compounds. Reaction rates tend to be sensitive to physical parameters like temperature, pH, and the presence of other molecules. Rate constants provide a clear link between reaction rates and these physical conditions, thereby allowing us to place the synthetic organic chemistry at the heart of the origin of life problem in a planetary context \cite{Lahav1982}.

The plausibility of prebiotic scenarios is often assessed from single values: a fixed temperature, pH, and reactant concentration yielding a singular product yield. Translating laboratory chemistry to natural environments requires moving beyond this, with kinetics \cite{White2024} informed by experimentally measured and computationally predicted rate constants playing a central role. These rate constants are themselves typically reported at fixed conditions \cite{Rimmer2021}, although much work has been done to explore the dependence of rate constants on pH \cite{White2024b}, buffer salts, and the nature of the participating nucleophiles and electrophiles \cite{Mayr2005}, a formidable challenge when water is the solvent of interest \cite{Brotzel2007}. 

Both experimental and theoretical determinations of rate constants carry inherent uncertainties, better represented as ranges than single values. These ranges reflect our true level of knowledge, and should be propagated through geochemical models of prebiotic environments. By quantifying uncertainty, we can better constrain the wider range of chemically possible outcomes and identify the quantities and conditions that would benefit from further experimental constraint. Critically, constraining the uncertainty in cyanide hydrolysis to changes in temperature, pH, and geochemically plausible salts will allow us to reliably constrain the maximum lifetime of hydrogen cyanide. Given a source flux of \ce{HCN}, we can provide robust constraints on the availability of cyanide in  the waters of Earth and other planets.

\section{Results}

We measured the hydrolysis rate of hydrogen cyanide as a function of temperature and pH, both in pure water and in the presence of high concentrations of hydrogen sulfide, sulfite, and phosphate. These elevated concentrations were chosen to maximize any effects on \ce{HCN} hydrolysis, not because they reflect prebiotically plausible conditions. We monitored formate as a proxy for \ce{HCN} hydrolysis: the immediate product, formamide, is comparatively short-lived and rarely observed, whereas formate is the ultimate hydrolysis product and its accumulation confirms that \ce{HCN} loss is due to hydrolysis rather than other processes.

\subsection{Hydrolysis of hydrogen cyanide in pure water}

We measured the concentration of formate as a function of time, temperature, and pH (see Supplementary Information). The concentration of formate was measured as a function of time using quantitative \ce{^1H} NMR (refer to Supplementary Figures S2--S170 for NMR spectra). Each measurement occupies a point in a five-dimensional data space defined by time, formate concentration, temperature, pH, and salt concentration (see Methods). The entire set of data is explained by a simple kinetic model based on acid and base catalyzed hydrolysis of \ce{HCN}. To find the most likely explanation for the data, in terms of hydrolysis rate constants, we employed a Bayesian statistical model.

We constrain the rates of acid-catalyzed and base-catalyzed HCN hydrolysis, expressed as:\\
\begin{equation}
\frac{\der [\ce{HCO2-}]}{\der t}=(k^{+}[\ce{H+}]+k^{-}[\ce{OH-}])\cdot{[\ce{HCN}]}
\label{eqn:rate-equation-diff}
\end{equation}
where $t$ (s) is time, $[\ce{H+}]$ (M) is the concentration of \ce{H+}, $[\ce{OH-}]$ (M) is the concentration of \ce{OH-}, $k^{+}$ (M$^{-1}$ s$^{-1}$) is the acid-catalyzed hydrolysis rate constant, and $k^{-}$ (M$^{-1}$ s$^{-1}$) is the base-catalyzed hydrolysis rate constant, $[\ce{HCO2-}]$ is the concentration of formate, and $[\ce{HCN}]$ the concentration of \ce{HCN}. Solving this equation and comparing it with measurements allows us to infer the distribution of likely rate constant values. These rate constants are cast as:
\begin{equation}
k^+ = k^{+}_{0} e^{-E_a^+/RT}, \;\;\;\;\;\;k^- = k^{-}_{0} e^{-E_a^-/RT},
\end{equation}
where $k^+, k^- \, ({\rm M^{-1} s^{-1}})$ are the rate constants for acid-catalyzed and base-catalyzed hydrolysis of \ce{HCN}, respectively, $k^{+}_{0}, k^{-}_{0} \; ({\rm M^{-1} \, s^{-1}})$ are the pre-factors for the rate constants, and $E_a^+, E_a^- \, ({\rm kJ \, mol^{-1}})$ are the activation energies. Our analysis provides the expected range of values for the prefactors $({\rm M^{-1} \, s^{-1}})$, $k^{+}_{0}$ and $k^{-}_{0}$ and the activation energies (kJ mol$^{-1}$), $E_a^+$ and $E_a^-$.

\begin{figure*}[t!]
\centering
\includegraphics[scale=0.5]{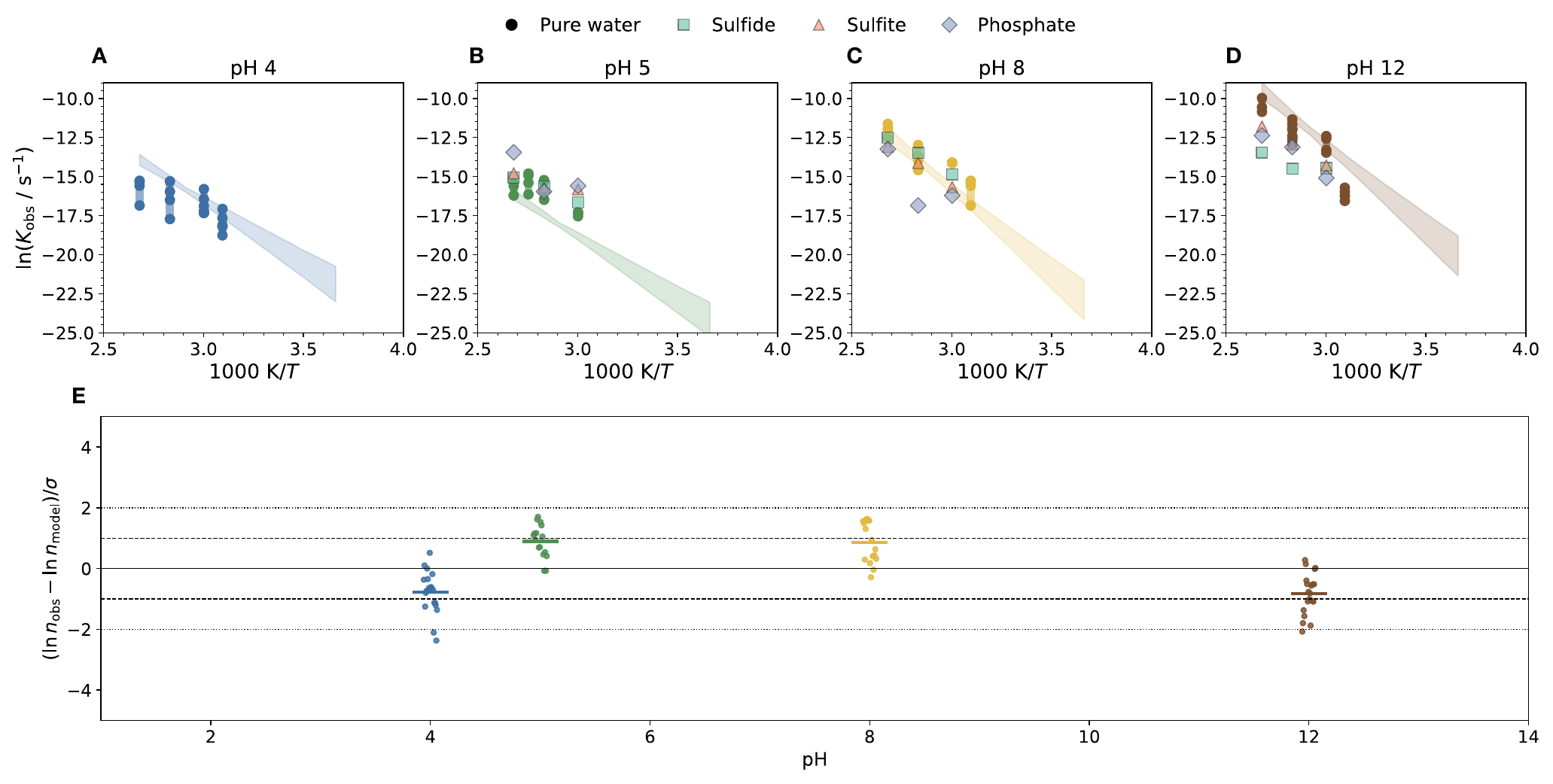}
\caption{Observed and best fit hydrolysis rates, as a function of temperature, with residuals. The top four panels show the observed rate of \ce{HCN} hydrolysis (y-axis) as a function of $1000 \, {\rm K}/ T$ (x-axis) at pH 4 (A), 5 (B), 8 (C), and 12 (D). The observed rate is defined here as the measured concentration of formate using NMR (see Methods and SI), divided by time. The shaded region represents the range of parameters that best explain the data, based on the range of best fit parameters $\{k_0^+,E_a^+,k_0^-,E_a^-\}$ (see Methods). Both pure water and salt rates are shown, but the fit shown is only the fit to the pure water data. Panel E shows the residuals from the fit, and reveal a weak dependence of the uncertainty on the pH.}
\label{fig:hydrolysis-fit}
\end{figure*}

Uncertainties in chemical rate constants follow a log-normal distribution (White et al.\ \textit{in prep}). We therefore express rate constants and their uncertainties as log-normal distributions, with temperature-dependence centered on the freezing temperature of water, $T^* = 273.15 \, {\rm K}$. Our best fits with uncertainties and residuals are shown in Figure \ref{fig:hydrolysis-fit}. We can express an effective rate constant that incorporates our best fits and correlated uncertainties using the following equation:
\begin{align}
\ln k_{{\rm 273}}^+ &= \ln k_{273}^+ - \dfrac{E_{a}^+}{R}\left(\dfrac{1}{T} - \dfrac{1}{T^*}\right), \notag\\
\ln k_{{\rm 273}}^- &= \ln k_{273}^- - \dfrac{E_{a}^-}{R}\left(\dfrac{1}{T} - \dfrac{1}{T^*}\right).
\end{align}
For convenient implementation in kinetic models, we provide a temperature-independent uncertainty represented by $\sigma_{\ln k^+},\sigma_{\ln k^-}$, corresponding to the uncertainty in the extrapolated rate constant at $273.15\ \mathrm{K}$. This is the largest uncertainty within the liquid-water temperature range and therefore provides a conservative approximation: it increasingly overestimates the uncertainty as the temperature approaches the range of our measurements. The full temperature-dependent uncertainty, including the covariance between $\ln k_{273}^+,\ln k_{273}^-$ and $E_{a}^+,E_{a}^-$, is described in the SI. Our resulting $\ln k_{273}^+$, $\ln k_{273}^-$, $E_a^+$, $E_a^-$, $\sigma_{\ln k^+}$, and $\sigma_{\ln k^-}$ are given in Table \ref{tab:hydrolysis-best-fit}.

\begin{table}[t!]
\centering
\caption{Parameters for the effective acid-catalyzed and base-catalyzed rate constants for \ce{HCN} hydrolysis \label{tab:hydrolysis-best-fit}}
\begin{tabular}{lSSSS}
\hline
Parameter & {No Salt} & {Sulfite} & {Sulfide} & {Phosphate} \\
\hline
$\ln[k_{273}^+/(1\,\mathrm{M}^{-1}\mathrm{s}^{-1})]$ & -12.7 & -10.0 & -10.8 & -9.8\\
$E_a^+ \; ({\rm kJ \, mol^{-1}}$) & 60.0 & 65 & 60.5 & 58.0\\
$\sigma_{\ln k^+}$ & 1.12 & 1.52 & 1.38 & 1.81\\
$\ln[k_{273}^-/(1\,\mathrm{M}^{-1}\mathrm{s}^{-1})]$ & -9.0 & -9.9 & -9.9 & -10.2\\
$E_a^- \; ({\rm kJ \, mol^{-1}})$ & 89.5 & 89.3 & 93.0 & 90.0\\
$\sigma_{\ln k^-}$ & 1.27 & 1.67 & 1.53 & 1.96\\
\hline
\end{tabular}
\end{table}

\subsection{Impact of sulfite, sulfide and phosphate on the hydrolysis of HCN}

We investigated the change in hydrolysis of \ce{HCN} in the presence of sulfite, sulfide and phosphate. The difference between the \ce{HCN} hydrolysis rate measured in pure water and in the presence of sulfite or sulfide salts likely arises from adduct formation. In the presence of sulfite, \ce{HCN} reacts to form a water-soluble adduct (Scheme \ref{sch:hydrolysis}). This reaction of sulfite with \ce{HCN} is a well-known reaction and its rate has been measured \cite{Rimmer2018}. The natural log of the rate of formation of the adduct, $\ln \big(k_{\rm AMDS}/1 \,{\rm M^{-1} \, s^{-1}}\big) =18.42-(79 \, {\rm kJ \, mol^{-1}}/RT)$, is comparable to the measured hydrolysis rate of \ce{HCN}. The presence of sulfite therefore has the potential to reduce the lifetime of \ce{HCN} beyond what would be predicted for pure water. Both the adduct and formate formation are faster at higher temperatures, suggesting that cooler water favors the persistence of dissolved \ce{HCN}. In the presence of sulfide, the formamide intermediate can form thioformamide (5) under both acidic and basic conditions (Scheme \ref{sch:hydrolysis}). The conversion of cyanide into thioformamide, and eventually thioformate, instead of formamide and formate results in less overall formate. As a result, the hydrolysis of cyanide appears to be slower under alkaline conditions.The formation of thioformamide  proceeds faster in the presence of thiols, and the presence of carbonyl impurities like acetone can indirectly accelerate the reaction \cite{Hyde2024}. Thioformamide is a useful intermediate for many important biomolecules like nucleobases and amino acids \cite{RN110,RN111}.
We predict that, if the thioformamide and thioformate concentrations were also accounted for, this would make up the difference in formate production. Such a project would be outside the scope of this work. We do not have a good mechanistic hypothesis for the accelerated rate of acid-catalyzed hydrolysis when sulfide is present. For the effect of sulfide on formate production, see Table \ref{tab:hydrolysis-best-fit}.

We observed a slower formate formation in the presence of phosphate. The discrepancy between hydrolysis with and without phosphate is greater at lower temperatures. At higher temperatures, hydrolysis rates with and without phosphate are comparable within the uncertainty previously determined. \ce{^1H} NMR data shows no peaks that could correspond to adducts (Figures S55-S93). Given the relevance of phosphate for prebiotic chemistry \cite{Miller1959}, further investigation on how phosphate salts can interfere with cyanide hydrolysis and other prebiotic reactions could be fruitful. 

\subsection{Modeling the Hydrolysis of Hydrogen Cyanide}

We modeled \ce{HCN} hydrolysis over 30 days to compare the predicted outcomes of a simplified prebiotic reaction network comparing literature values from Miyakawa \textit{et al.}\ \cite{Miyakawa2002} with those derived in this work. The chemical reactions included in the model, along with their rate constants and literature sources, can be found in Table S1.

\begin{figure*}
    \centering
    \includegraphics[width=0.8\linewidth]{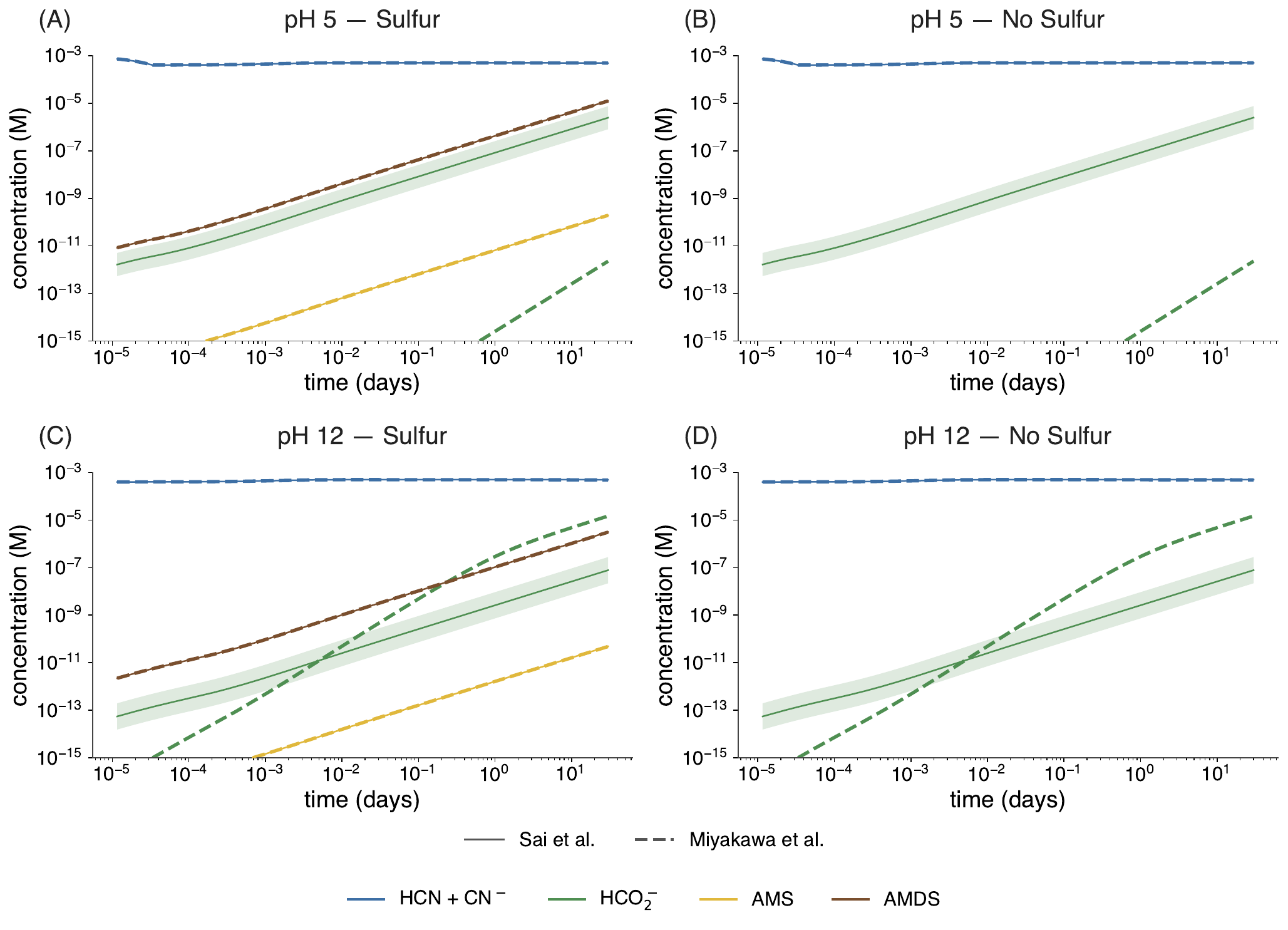}
    \caption{Hydrogen cyanide hydrolysis in a prebiotic reaction network. We model the reaction network detailed in Table S1 over 30 days using the SOUP model of White \textit{et al.}\ (\textit{in prep}), at pH 5 and 12, with and without sulfur. Dashed lines use the hydrolysis rate constants of Miyakawa \textit{et al.}\ \cite{Miyakawa2002}; solid lines use the rate constants determined in this work. Colours indicate species: blue, \ce{HCN} + \ce{CN-}; green, \ce{HCO2-}, where the shaded band shows the uncertainty obtained by propagating $\pm\sigma$ on $\ln k^+$ and $\ln k^-$; yellow, aminomethanesulfonate; brown, aminomethanedisulfonate.}
    \label{fig:soup}
\end{figure*}

Taking this reaction network we use the Survivability Of prebiotic chemistry Under Planetary conditions (SOUP) framework outlined by White \textit{et al.}\ (\textit{in prep}) to model the progression of this chemistry. We set the initial starting conditions in our aqueous system to include 1 mM of hydrogen cyanide as an optimistic estimate based on the maximal amount that would be delivered by impacts (Anslow \textit{et al.} \textit{accepted}). In addition, we include 100 $\mu$M of Fe(II) which we find to be a reasonable intermediate value from the literature \cite{Catling2020, Holland2020}. These levels are consistent with the anoxic conditions of the Hadean, under which the absence of atmospheric oxygen suppresses the ferrous-to-ferric transition and allows Fe(II) to accumulate. We also include 1 mM of sulfur dioxide speciated as \ch{HSO3-} and \ch{SO3^{2-}} in line with Ranjan \textit{et al.}\ who state that during episodes of intense volcanism \ch{SO2}-derived anions may be available at mM levels \cite{Ranjan2018}. We note that this is specific to reservoirs buffered to a pH of approximately 7 making our estimate an optimistic upper bound. Finally, we set the pH to be buffered at 5 and 12, displaying results for each of these cases in Figure \ref{fig:soup}. 

Our results show a two-phase depletion of cyanide. At both pH conditions, Fe(II) is exhausted quickly locking up some of the cyanide in ferro-cyanide complexes. The amount of Fe(II) is insufficient to sequester all of the cyanide, leaving the remaining cyanide susceptible to hydrolysis. This is reflected in the gradual accumulation of formate whose concentration continues to increase throughout the 30-day simulation in all panels.

Comparing the rate constants derived in this work with those of Miyakawa \textit{et al.}\ \cite{Miyakawa2002} reveals a chemically significant reversal between the two hydrolysis regimes. At pH 12, where hydroxide concentrations favor base-catalyzed hydrolysis, the Miyakawa \textit{et al.}\ rate constants predict greater formate production at the end of the 30-day period. In contrast, at pH 5, where acid-catalyzed hydrolysis dominates, the relationship reverses and the rate constants derived here predict substantially more formate. The formate production predicted by the SOUP model shows a difference at ${\rm 25^{\circ}C}$ over 10 days at pH 5 that exceeds five orders of magnitude. Even at pH 12, where the rate constants from Miyakawa \textit{et al.} and this work are much closer, the predicted formate yield at ${\rm 25^{\circ}C}$ after 10 days still differs by more than an order of magnitude. These results highlight how rate constants that are similar at high temperatures can diverge substantially when extrapolated to lower temperatures.

We observe that the inclusion of sulfur species has very little impact on the chemical outcome. The sulfonation products aminomethanesulfonate (AMS) and aminomethanedisulfonate (AMDS) show a weak sensitivity to both reaction network and pH. Because this reaction is limited by the availability of its substrate, which is only weakly depleted over the 30-day simulation regardless of the hydrolysis rate constants used, AMS and AMDS yields are essentially independent of the formate formation rates. The apparent pH-insensitivity is more mechanistically notable: the AMS and AMDS rate laws contain no explicit pH terms, meaning pH enters only indirectly via the two included acid-base equilibria that respond in opposite directions to a change in pH, lowering pH increases [\ce{HCN}] but proportionally decreases [\ce{SO3^{2-}}].

We note that the simulations described above were only run for 30 days; we anticipate reaction outcomes would diverge more substantially over longer timescales, however the characteristic timescales over which meaningful deviations would emerge, years to centuries depending on pH and mechanism, rendered longer integrations computationally prohibitive.

\section{Discussion}
We used measurements of cyanide hydrolysis rates to constrain the acid-catalyzed and base-catalyzed rate constants for cyanide in pure water and measured how the rates are affected by the presence of a variety of geochemically plausible salts. In the absence of salts, we find that our measured hydrolysis rate constants agree well with previously reported base-catalyzed values in the literature, but diverge significantly for the acid-catalyzed values. Interestingly, we find that model predictions for formate formation using the Miyakawa et al.\ rate constants differ significantly from the predictions using our rate constants. We find that salts slow down the rate of hydrolysis in alkaline conditions and speed up the rate of hydrolysis in acidic conditions. In addition, the presence of salts introduces other cyanide destruction pathways.

\begin{figure}[t!]
\centering
\includegraphics[scale=0.55]{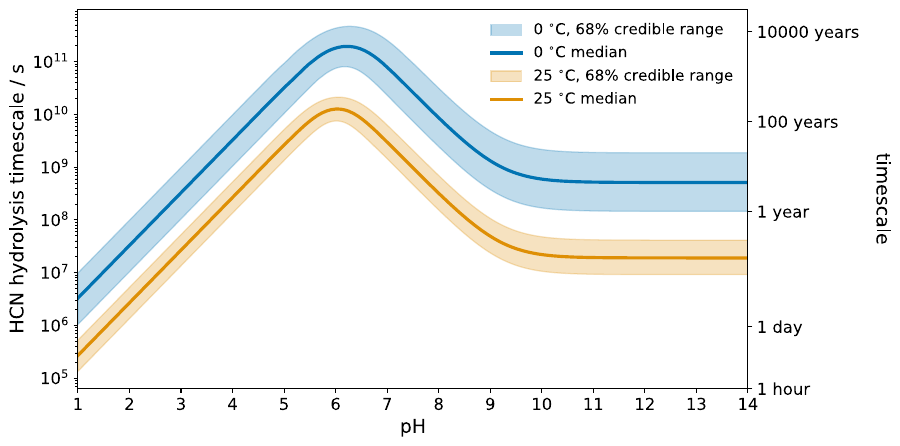}
\caption{The lifetime of \ce{HCN} as a function of pH, for ${\rm 0^{\circ}C}$ and ${\rm 25^{\circ}C}$ with uncertainty. The longest lifetime for \ce{HCN} is found to be near pH 6--7. The peak lifetime shifts slightly to lower pH as temperature increases. The uncertainty implies predicted lifetimes of \ce{HCN} at low temperatures can differ by an order of magnitude, and that predictions about the pH that correspond to a specific lifetime can differ by $\pm 1$.}
\label{fig:hcn-lifetime}
\end{figure}

 We use the hydrolysis rate constants obtained in this work to estimate the lifetime of cyanide in water at two different temperatures, ${\rm 0^{\circ}C}$ and ${\rm 25^{\circ}C}$ (Figure \ref{fig:hcn-lifetime}). We see that \ce{HCN} survives longest at low temperatures and neutral pH. At ${\rm 0^{\circ}C}$ and at a pH between 6 and 7, \ce{HCN} has a lifetime of over 1000 years. At a pH $\gtrsim 9$, the lifetime falls to between 1 and 50 years. And at a pH $\lesssim 4$, the lifetime is less than a year. 
 
\begin{figure}[t!]
\centering
\includegraphics[scale=0.55]{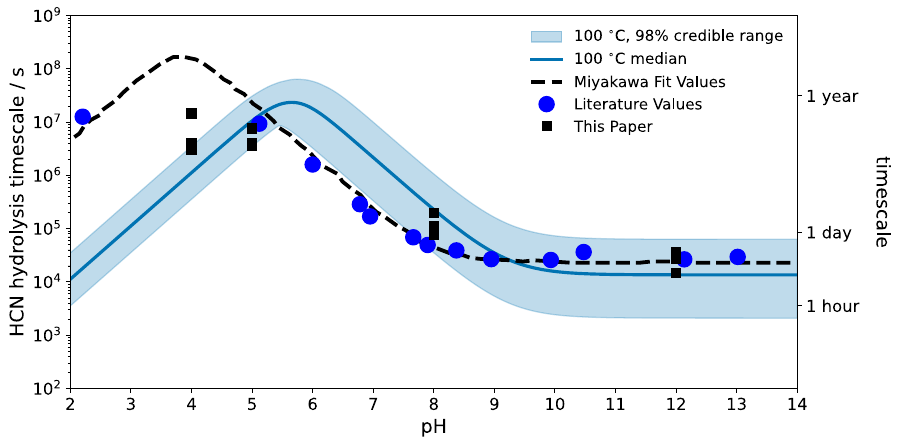}
\caption{The lifetime of \ce{HCN} as a function of pH, for ${\rm 100^{\circ}C}$ with uncertainty. Comparison between our results and literature values, taken from Miyakawa \textit{et al.} \cite{Miyakawa2002}. Our results are consistent with Miyakawa down to pH $\sim 5$, after which they diverge significantly.}
\label{fig:hcn-lifetime-compare}
\end{figure}

 We compare our results at ${\rm 100^{\circ}C}$ to those of Miyakawa \textit{et al.} and references therein \cite{Miyakawa2002}. Consistent with Miyakawa \textit{et al.}, we find that cyanide hydrolyses more slowly at lower temperatures. However, our range of findings make the ideal pH for cyanide survival broader and more neutral, suggesting maximal lifetimes between pH 6--7 rather than Miyakawa \textit{et al.}'s pH 4 (see Figure \ref{fig:hcn-lifetime-compare}). When results that appear close at ${\rm 100^{\circ}C}$ are extrapolated down to lower temperatures, such as ${\rm 25^{\circ}C}$ or ${\rm 0^{\circ}C}$, divergences grow substantially.

We note that none of our results rule out the possibility of productive prebiotic chemistry with hydrogen cyanide at any pH. They simply indicate that cyanide concentrations are lower outside the pH 6--8 range, and that any excess cyanide introduced under these conditions is removed more rapidly. These results are especially powerful when applied to aqueous chemical kinetics models as was demonstrated in Figure \ref{fig:soup}. Here we find that the choice of rate constant strongly impacts the amount of formate that can accumulate in prebiotic environments. Although we do not include sink terms for formate, the relative differences in yield between rate constant choices are expected to be robust, since any destruction pathway would act equally on both networks and therefore preserve the order-of-magnitude differences in accumulation reported here.

\subsection{Prebiotic Scenarios that Require Cyanide}

We can explore the implications of our cyanide hydrolysis rate constants for several proposed sources of \ce{HCN} in the Hadean Earth environment ($\geq 4 \, {\rm Gya}$). Most of these involve forming \ce{HCN} in the atmosphere, where it then rains out onto the surface. Pearce \textit{et al.} give the current best constraints for cyanide production in the Hadean atmosphere \cite{Pearce2022}. There are several planetary sources of cyanide we could consider, as discussed in our introduction, from cometary delivery to lightning generation in a reducing atmosphere. Here, we will consider three:

\begin{enumerate}
	\item Predicted photochemical production of \ce{HCN}  and rainout from a volcanically-derived Hadean atmosphere \cite{Gaillard2014,Tian2011}. Here the \ce{HCN} flux is $10^5 \; {\rm cm^{-2} \, s^{-1}}$ 
	\item Predicted photochemical production of \ce{HCN} and rainout from a post-impact Hadean atmosphere \cite{Zahnle2020,Wogan2023}. Here the \ce{HCN} flux is $10^8 \; {\rm cm^{-2} \, s^{-1}}$.
	\item Predicted flux into a surface hydrothermal system from an ultra-reduced magmatic gas \cite{Rimmer2019a,Rimmer2024}. Here the \ce{HCN} flux is $10^{10} \; {\rm cm^{-2} \, s^{-1}}$.
\end{enumerate}

\begin{figure}[t!]
\centering
\includegraphics[scale=0.55]{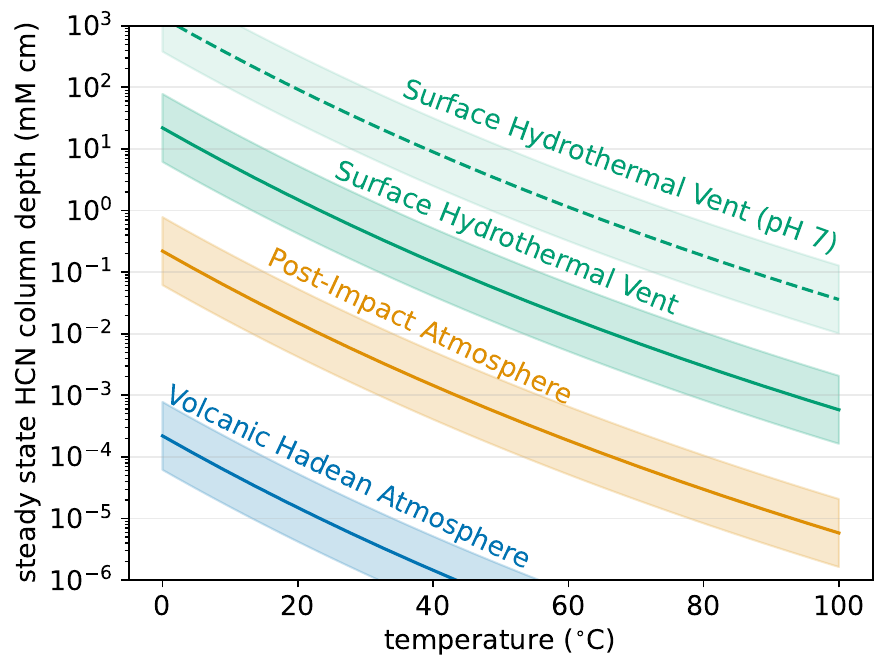}
\caption{Estimated steady state column depth of \ce{HCN} (with units of mM cm) in an alkaline lake (pH 9) as a function of temperature for photochemical \ce{HCN} rainout from a volcanic Hadean atmosphere ($\Phi = 10^5 \; {\rm cm^{-2} \, s^{-1}}$), from a post-impact atmosphere ($\Phi = 10^8 \; {\rm cm^{-2} \, s^{-1}}$), and a surface hydrothermal vent sourced by reduced komatiitic magmatic gas ($\Phi = 10^{10} \; {\rm cm^{-2} \, s^{-1}}$). We also show the same surface hydrothermal scenario but with the lake at a pH of 7. The column can be divided by the average depth to give an average concentration.}
\label{fig:lake-model}
\end{figure}

We will consider a common environment for these three cyanide source scenarios: an alkaline lake of variable depth at pH 9 and a range of temperatures. We apply the above-listed fluxes and set them against hydrolysis for pure water (see Methods) to estimate the steady-state column of \ce{HCN} in the lake due to these different sources. The column is represented as the concentration times the average depth of the body of water, with units ${\rm mM \, cm}$. One simply divides by the average depth to find the average concentration of \ce{HCN}. Our estimates of the column are given as a function of temperature in Figure \ref{fig:lake-model}.

We employ an alkaline lake for these models because some alkaline lakes would likely be rich in dissolved phosphate \cite{Catling2024}, and could store some of the dissolved hydrogen cyanide as cyanide salts \cite{Toner2019, Patel2015}. Also, some prebiotic chemistry has been tested and found to work well in these environments \cite{Cohen2024}. These environments are predicted to range in pH from 7 to 10. Though these lakes have high concentrations of salts, on early Earth they would have been buffered by large partial pressures of carbon dioxide, pulling their pH down to 7 \cite{Toner2019}. We next examine how different atmospheric sources of \ce{HCN} translate into dissolved cyanide concentrations in these waters.

A volcanic-gas-derived Hadean atmosphere will be dominated by carbon dioxide and molecular nitrogen \cite{Zahnle}, with up to 5\% \ce{H2} \cite{Zahnle2019}. Hydrogen cyanide is not produced in great abundance in the volcanic Hadean atmosphere, owing to its unfavorable redox conditions, especially its low C/O ratio \cite{Rimmer2019a}. Nevertheless, some models suggest a considerable rainout flux of \ce{HCN} of up to $10^5 \; {\rm cm^{-2} \, s^{-1}}$ \cite{Tian2011}. With that rainout flux, there are small amounts of dissolved \ce{HCN} at steady state in an alkaline lake, lower than $1 \, {\rm \mu M \, cm}$, even when the water is near freezing.

Post-impact atmospheres, though transient, could be far more reducing, if the iron carried by the impactor reacts with water, oxidizing the iron to release \ce{H2} \cite{Genda2017}. The \ce{H2} eventually escapes, and the oxidation state returns to its pre-impact levels \cite{Zahnle2020}. Photochemical \ce{HCN} production is far more favorable in these environments, resulting in fluxes a thousand times that of the most favorable volcanic-atmosphere case \cite{Wogan2023}. With this enhanced rainout into an alkaline lake the situation looks far more similar to that encountered in a prebiotic chemistry experiment, with a column approaching ${\rm 1 \, mM \, cm}$. If the lake is buffered by excess \ce{CO2} or \ce{SO2}, the pH will drop further, and could rest around 7, in which case columns of $10 \; {\rm mM \, cm}$ or even $100 \; {\rm mM \, cm}$ are possible.

Hadean magmas were plausibly hotter than modern magmas, and organic material delivered or produced during post-impact atmospheric chemistry could have been incorporated into magma and converted to graphite. A thin gas-phase is held in equilibrium with graphite, which models predict will result in even more \ce{HCN}, along with other nitriles \cite{Rimmer2024}. With this flux of \ce{HCN} bubbling into an alkaline lake, the lake's \ce{HCN} column can reach $10 \; {\rm mM \, cm}$, or even up to nearly $1 \; {\rm M \, cm}$ if the lake were at pH 7. 

These scenarios are all Earth-centered, but they can help inform how prebiotic chemistry may play out on other planets.

\subsection{Prebiotic Synthesis on Other Planets: Further Connecting the Abiogenesis and Habitable Zones}

Future work can combine our cyanide hydrolysis rate constants with models of atmospheric production, rainout, and aqueous accumulation \cite{Todd2024} to estimate the abundance and persistence of cyanide in the surface waters of other worlds. Atmospheric observations may constrain the abundance of \ce{HCN} and related nitriles \cite{Friss2026}, while climate models can place bounds on surface temperature. Measurements or models of atmospheric \ce{CO2}, \ce{SO2}, and other acid-forming gases, together with the expected mineralogy of the surface, can further constrain the likely pH of exposed waters. Our rate constants then provide the link between these planetary conditions and the lifetime of dissolved cyanide.

This framework could be used to translate a future detection of atmospheric \ce{HCN} on a rocky planet into estimates of the amount of cyanide available in the ocean, lakes, ponds, or other open waters, including the fraction stored as iron-cyanide complexes and the rate at which hydrolysis produces ammonium formate. When interpreted together with inderect constraints on surface temperature and aqueous pH, the observation of gas-phase \ce{HCN} could provide a quantitative constraint on prebiotic chemical inventories at the planetary surface.

Planets with lower surface temperatures favor the persistence of cyanide, potentially making cool planets near the outer edge of the habitable zone especially favorable for its accumulation. Similar arguments may also apply to colder worlds with more unusual atmospheric structures, including proposed Hycean planets \cite{Madhu2021}. Future observations with JWST and other facilities may therefore allow atmospheric measurements to be connected directly to the chemical conditions required for prebiotic synthesis \cite{Giacobbe2021,Claringbold2023}.

The rate constants we have measured can have wide-spread application to atmospheric and surface chemistry models and experiments. Often, first approaches for connecting atmospheric chemistry to aqueous environments consider rainout as a source of molecules to the water column, and neglect any appreciable sink \cite{Pinto1980}. This picture of plausibility, however, can change dramatically when sinks are included \cite{Scherf2026}. This is even more pronounced when uncertainty is included and ranges of scenario-specific outcomes consistent with experimental uncertainty are considered.

\section{Conclusion}
We experimentally determined the hydrolysis rate of \ce{HCN} with associated uncertainties, over a range of temperatures, pH values, and in the presence of phosphate, sulfite and sulfide. Under alkaline conditions, we observed that the activation energy for cyanide hydrolysis is comparable to literature values when accounting for uncertainty. However, under acidic conditions, \ce{HCN} hydrolysis rates have an activation energy that is significantly lower, and so the rates are significantly faster. The introduction of salts in general is observed to slow down hydrolysis under alkaline conditions and speed up hydrolysis under acidic conditions. Particularly, the rates are 2.5 times slower at ${\rm pH \gtrsim 8}$ in the presence of sulfite, and sulfide. We show that \ce{HCN} can survive for up to thousands of years at near freezing temperatures in a neutral environment and in doing so demonstrate that constraining hydrolysis rate constants, and aqueous reaction kinetics more broadly, is essential for distinguishing which prebiotic scenarios are chemically plausible.

\section*{Materials and Methods}

\begingroup
\small

\subsection*{Materials and Characterisation}
All the chemicals used for the study, \ce{KCN} (CAS no. 151-50-8), \ce{D2O} (CAS no.7789-20-0), sodium bicarbonate (CAS no.144-55-8), sodium sulfite (CAS no.144-55-8), sodium hydrosulfide hydrate (CAS no. 207683-19-0 ), sodium phosphate monobasic (CAS no. 7558-80-7), were purchased from Sigma Aldrich.  

Nuclear magnetic resonance spectroscopy (NMR) was used for characterisation of the reaction mixture. The data was recorded on an AVIII 500 MHz Spectrometer, equipped with a \ce{^1H} ``dual'' helium-cooled cryoprobe, and operated under the Microsoft Windows operating system running Topspin 3.7. The temperature of the probe was consistently maintained at $298 \, {\rm K}$ for each measurement. Chemical shifts for \ce{^1H} NMR spectra are reported as $\delta$ in units of parts per million (ppm) downfield from \ce{SiMe4} ($\delta$ 0.0). 

\subsection*{Sample preparation}
A solution of \ce{KCN} ($23 \, {\rm mM}$) and potassium bicarbonate ($71 \, {\rm mM}$) was prepared in an Argon-degassed 20\% \ce{D2O}/\ce{H2O} mixture. The pH of the mixture was adjusted using degassed hydrochloric acid ($12 \, {\rm M}$) and sodium hydroxide (50\% in \ce{H2O}). The solution was then heated in the presence or absence of a variety of salts, each introduced at $50 \, {\rm mM}$ concentrations, to study the stability of \ce{HCN} under different physical conditions. 

\subsection*{Estimation of concentration of cyanide over time}
For quantitative \ce{^1H} NMR analysis, aliquots of $400 \, {\rm \mu L}$ were taken and placed in an NMR tube with an internal standard ($50 \, {\rm mM}$ pentaerythritol ($4\,{\rm \mu L}$)). The concentration of formate was measured using an internal standard. This technique indirectly measures the hydrolysis of cyanide over time, and therefore is valid only when the cyanide hydrolyses completely to formate i.e., the concentration of the intermediate formamide is small. The concentration of formate $[\ce{HCO_2^-}]$ can be expressed as:
\begin{equation}
[\ce{HCO_2^-}] = \dfrac{I(\ce{HCO_2^-})} {I(\ce{PET})}\times [\ce{PET}]
\label{eqn:NMR-formate}
\end{equation}
where $I(\ce{HCO_2^-})$ is the integration area of the formate signal per hydrogen,  $[\ce{PET}]$ is the concentration of pentaerythritol standard and $I(\ce{PET})$ is the integration area of the pentaerythritol standard signal per hydrogen. By using equation (\ref{eqn:NMR-formate}) the concentration of the product could be tracked over time. 

\subsection*{Determination of Uncertainties}

To determine uncertainties on the acid- and base-catalyzed rate constants, a forward model was constructed to predict the formate concentration as a function of time. Predicted concentrations were obtained by integrating the rate equation (Equation~\ref{eqn:rate-equation-diff}), yielding:
\begin{equation}
\ce[\ce{HCO_2^-}] = Y[\ce{KCN}]_0\left(1 - e^{-kt}\right)
\label{eqn:model}
\end{equation}
where $Y \leq 1$ is the yield of cyanide hydrolysis, $[\ce{KCN}]_0 \ ({\rm M})$ is the initial amount of potassium cyanide included in the experiment, and $k$ is the effective rate constant, defined as:
\begin{equation}
k = k^{+}[\ce{H+}] + k^{-}\left(\frac{[\ce{OH-}]}{1 + 10^{(\mathrm{pH} - \mathrm{p}K_a)}}\right),
\label{eqn:rate-constant}
\end{equation}
where $\pKa$ is the dissociation constant of \ce{HCN}, and both $\mathrm{pH}$ and $\pKa$ are temperature dependent \cite{Bandura2006,Arcis2024,Verhoeven1990}. Both rate constants were assumed to follow Arrhenius temperature dependence:
\begin{equation}
\begin{split}
\ln k^{+} &= \ln k^{+}_0 - \frac{E_a^{+}}{RT},\\
\ln k^{-} &= \ln k^{-}_0 - \frac{E_a^{-}}{RT}
\end{split}
\end{equation}
The model was therefore parametrised by the set:
\begin{equation}
\theta = {k^+_0, E_a^+, k^-_0, E_a^-}
\end{equation}

\subsubsection*{Prior Distributions}
Uniform priors were assigned to each parameter in log-space, with bounds chosen to be sufficiently broad to encompass all physically plausible values. 

\subsubsection*{Likelihood Function}
The posterior probability of a parameter set $\theta$ given the data $D$ was evaluated using Bayes' theorem:
\begin{equation}
P(\theta \mid D) \propto \mathcal{L}(D \mid \theta), P(\theta)
\label{eqn:posterior}
\end{equation}
where $P(\theta)$  is the prior described above and the normalization constant $P(D)$ cancels when comparing relative probabilities across the parameter grid. A Gaussian likelihood function was adopted, such that:
\begin{align}
\ln \mathcal{L} =& -\frac{1}{2}\sum_{t,T,\mathrm{pH}} \dfrac{1}{\sigma}\big([\ce{HCO2-}]_{\mathrm{exp}}(t,T,\mathrm{pH}) \notag\\
&- [\ce{HCO2-}]_{\mathrm{th}}(t,T,\mathrm{pH}\mid\theta)\big)^2 \notag\ \\
&- \sum_{t,T,\mathrm{pH}} \ln\sigma - \frac{N}{2}\ln(2\pi)
\end{align}
where $[\ce{HCO2-}]_{\mathrm{exp}}$ and $[\ce{HCO2-}]_{\mathrm{th}}$ are the experimentally measured and model-predicted formate concentrations at a given time $t$, temperature $T$, and pH, and $N$ is the total number of data points. The per-point uncertainty $\sigma$ was treated as a free parameter, determined by maximizing $\mathcal{L}$ at each point in the parameter grid. This allowed each datapoint to inform on the uncertainty, but because not all experiments were repeated many times under the same conditions, this method can generate systematics in temperature and pH. The systematics generated in temperature were found to be negligible, and the systematics in pH significant, but within $\pm \sigma$ at the best $\sigma$ value we find, see Figure \ref{fig:hydrolysis-fit}.

\subsubsection*{Parameter Space Exploration}
The posterior was explored using an iterative grid search strategy. An initial coarse grid was constructed spanning the full prior range for each parameter in $\theta$. The region of parameter space yielding the highest posterior probabilities was identified, and the search was repeated twice at progressively finer resolution, each time re-centering the grid on the best-fitting region. The resulting posterior distributions over $\theta$ were used to derive best-fit values and uncertainties for the acid- and base-catalysed rate constants. Fits and residuals are shown in Figure~\ref{fig:hydrolysis-fit}. The posterior probabilities can be seen in the SI.

\subsection{The Lifetime of HCN}
The rate constants obtained in this work were applied to Equation (\ref{eqn:rate-constant}) to find the lifetime of \ce{HCN}: 
\begin{equation}
\tau_{1/2}=\frac{\ln 2}{k},
\end{equation}
These were then used to produce Figures \ref{fig:hcn-lifetime} and \ref{fig:hcn-lifetime-compare}. 

Finally, if an input flux of \ce{HCN}, $\Phi(\ce{HCN})$, with units ${\rm cm^{-2} \, s^{-1}}$, were introduced into a body of water of known depth, $\ell \, ({\rm cm})$, then the steady state concentration of \ce{HCN} would be set by:
\begin{equation}
\dfrac{\der [\ce{HCN}]}{\der t} = \big(1.66 \times 10^{-21} \; {\rm M \, cm^3}\big)\dfrac{\Phi(\ce{HCN})}{\ell} - k [\ce{HCN}] = 0.
\end{equation}
and the column, $[\ce{HCN}] \ell$ can be determined, as shown in Figure \ref{fig:lake-model}.

\subsection*{SOUP Model of Cyanide Hydrolysis}

The progression of a prebiotic reaction network was modeled using the Survivability Of prebiotic chemistry Under Planetary conditions (SOUP) framework described by White \textit{et al.}\ \textit{in prep}. This is an aqueous chemical kinetics model that simulates coupled aqueous reaction kinetics and gas exchange by solving a system of ordinary differential equations (ODEs). For a full description of the model see White \textit{et al.}\textit{in prep}. 

The reaction network used in this work was adapted from White \textit{et al.}\ \textit{in prep} to include the rate constants determined in this study. See Supplementary Information for the full network. The model was used to simulate this reaction network over 30 days at pH 5 and pH 12, both with and without sulfur chemistry.

The conditions of the model are $T = {\rm 25^{\circ}C}$, ${\rm pH = 5, 12}$, $[\ce{HCN}]_0 = {\rm 1 \, mM}$, $[\ce{Fe^{2+}}] = 0.1 \, {\rm mM}$, and  $[\ce{SO2}]_0 = {\rm 0 \, mM, 1 \, mM}$ for the cases without and with sulfur, respectively. All of these are dissolved species. Here we do not model air-water exchange of molecules.

\endgroup

\dataavail{All the datasets analyzed in this study and the associated modeling data are available in the Harvard Dataverse repository: 
\href{https://dataverse.harvard.edu/dataset.xhtml?persistentId=doi:10.7910/DVN/WT6NKZ}{doi:10.7910/DVN/WT6NKZ}}.

\acknow{S.S.M. thanks Duncan Howe and his team from the NMR facility of the Yusuf Hamied Department of Chemistry, and David Russell from Biochemistry Department, both at University of Cambridge, for help with the \ce{^1H} NMR. We thank the Leverhulme Centre for Life in the Universe Joint Collaborations Research Project Grant G112026, Project KKZA/237, and the Royal Society Grant G125370 for funding this project. S.B.W was supported by Newnham College, the Leverhulme Centre for Life in the Universe (G112026, Seed Fund, Project KKZA/253), and the Department of Physics.}

\showacknow{} 

\bibsplit[32]


\end{document}